\documentstyle[epsf,preprint,eqsecnum,aps]{revtex}

\begin{document}

\preprint{\vbox{\hbox to \hsize{\hfill UTREL 95-02}
	           \break\nointerlineskip\vskip 4pt
                \hbox to \hsize{\hfill hep-lat/9502008}}}

% -- \bibliographystyle{prsty}
\draft

\title{Failure of Universality in Noncompact Lattice Field Theories}

\author{T. E. Gallivan\footnote{timg@landau.ph.utexas.edu}
                and Arie Kapulkin\footnote{arik@landau.ph.utexas.edu}}

\address{
        Center for Relativity \\
        Department of Physics \\
        University of Texas at Austin \\
        Austin, Texas 78712-1081
}

\date{\today}
\maketitle

\begin{abstract}

The nonuniversal behavior of two noncompact nonlinear sigma models is
described. When these theories are defined on a lattice, the behavior
of the order parameter (magnetization) near the critical point is
sensitive to the details of the lattice definition. This is counter to
experience and to expectations based on the ideas of universality.

\end{abstract}


%% Just before electronic submission comment out the \input secX
%% commands and insert the appropriate files into the master file

%
%       Section 1
%

\section{Introduction}
\label{intro}

Functional integration is an extremely useful tool for the
non-perturbative analysis of quantum field theories (QFT from now on).
To ensure that calculations involving these integrals be well defined,
it is convenient to regularize them by working on a spacetime lattice.
A continuum QFT is then {\em defined\/} in terms of a sequence of
lattice field theories.  There may be many lattice representations for a
single continuum theory, so it is not obvious that such a definition is
unique.  In the majority of lattice field theory simulations, the issue
of uniqueness is addressed by appealing to the principle of
universality.  Universality is the property that, in the continuum
limit, the system ceases to depend on the exact nature of the lattice
theory, and many different lattice field theories lead to the same QFT.
In the approach to the continuum limit, which is located at
special---critical---values of the coupling parameters, various
quantities are expected to have a universal dependence on the couplings.
Two different lattice theories that display the same behavior at this
critical point are said to belong to the same universality class.

In the present paper, lattice models that do not to conform to this
expectation of universality are described, and the extent of universal
behavior is clarified somewhat. Four models are considered here: two
lattice versions of the $SO(1,1)$ nonlinear sigma model and two lattice
versions of the $SO(2,1)$ sigma model.  In each case the two lattice
models become the same classical theory in the {\it classical\/}
continuum limit.

Numerical evidence strongly suggests \cite{DDF:92a,DDF:92b,DDF:88} that
both lattice $O(2,1)$ models {\em do\/} belong to the same universality
class in the sense that they both describe a pair of free uncoupled
massless scalar fields in the continuum limit (they are trivial).  The
situation for the $SO(1,1)$ models is not as clear. One lattice
$SO(1,1)$ model is known to become a free massless scalar field in the
continuum limit, but the continuum limit of the other lattice model is
unknown. The current numerical data does not rule out triviality.

All of these lattice theories, however, have an order parameter, called
the magnetization, which {\em is} sensitive to the details of the
lattice model near the critical point.  This is unlike the behavior
observed in more familiar models, such as the Ising model or the {\it
compact\/} nonlinear sigma models \cite{Ami:84}. It is argued that the
cause for this nonuniversal behavior is the radically different scaling
of the fields with a parameter $\beta$ (analogous to an inverse
temperature in statistical mechanics) that is made possible by the
noncompact configuration spaces. This point will be further discussed in
Section \ref{concl}.

The remainder of the paper is organized as follows.  In Sec.
\ref{univ}, the nature of the approximation involved in defining a
QFT on a lattice is explained, and the freedom involved in this
approximation is demonstrated. The relevance of universality to lattice
field theory is then explained. The language of the renormalization
group (RG from now on) is used to show why certain quantities ({\it
e.g.\/} critical exponents) are expected to be the same for lattice
field theories that are in the same universality class. In addition, the
process through which two different lattice theories become the same QFT
in the continuum limit is explained in terms of the concept of
irrelevant operators. The general structure of nonlinear sigma models is
given in Sec. \ref{models}. Sec. \ref{o11} introduces two possible
lattice $SO(1,1)$ models and shows that they have different behavior in
the critical region. Sec. \ref{o21} discusses the analogous situation
for the $SO(2,1)$ model. In Sec.
\ref{concl} the reason for the nonuniversal behavior is discussed, and
the conclusions are presented.

It should be noted that unlike the two and three dimensional models
that are of interest in the related discipline of critical phenomena,
this article deals with models in four spacetime dimensions, since the
motivation is to study quantum field theory. To carry out numerical
simulations of the lattice field theory, it is necessary to analytically
continue the time into the Euclidean regime. The result of this
Euclideanization is mathematically equivalent to a statistical
mechanical partition function, with a Hamiltonian given by ${H\over kT}
= S_E$ where $S_E$ is the Euclidean action of the lattice field theory.

%%% -- End of first section
%
%       Section 1a
%

\section{Lattice Field Theories and Universality}
\label{univ}

A quantum field theory is constructed from a lattice field theory as
follows. The QFT is defined by the vacuum expectation values of time
ordered products of operators built out of the local quantum fields.
These time ordered products can be expressed in terms of functional
integrals. The functional integrals are in turn defined by discretizing
spacetime, which turns them into multiple integrals, albeit in a very
high dimensional space. The multiple integrals are then amenable to
numerical analysis through Monte Carlo techniques. The lattice field
theory approximants to the expectation values of the continuum QFT can
then be calculated\footnote{Of course, non-numerical techniques may also
be used.}. The simulations are carried out on successively larger
lattices, and the approximants are then extrapolated to the continuum
limit to yield the desired continuum vacuum expectation values.

Thus, there are two identifiable stages in the process of quantization
through functional integrals. First, a lattice field theory is defined
as a large dimensional multiple integral with an action constructed from
the classical action. Second, the continuum QFT is recovered as the
limit of a sequence of lattice theories simulated on successively larger
lattices, with the lattice spacing simultaneously considered to be taken
to zero.  If the continuum QFT contains any finite length scale, such as
the inverse of the mass or the correlation length $\xi$, this length
scale becomes infinite when measured in units of the lattice spacing
$a$, which goes to zero in the continuum limit.  Therefore, in the
second stage, when the lattice spacing is taken to zero, it is necessary
to identify a point in the parameter space of the theory where the
lattice version of the characteristic length scale of the QFT diverges
in units of the lattice spacing. The correlation length of the nonlinear
sigma models discussed here is already infinite on an infinite lattice
because the models are massless. Instead, the relevant length scale here is
 $\mu_R^{-1}$ which is associated with the interaction. The
continuum limit is sought where the characteristic length scale measured
in units of the lattice spacing, ${1\over \sqrt{\beta_R}}$, diverges
($\beta_R\equiv\mu_R^2 a^2$). This means that a continuum limit exists
only where $\beta_R\rightarrow 0$.

The lattice action constructed during the first stage is not unique,
because the continuum action contains derivatives of the fields.  The
customary prescription for passing from the classical continuum action
to the lattice action is to replace $\partial_\mu\phi(x) \rightarrow
\frac{1}{a}\Delta_\mu\phi(x)$, where $\Delta_\mu\phi(x) = \phi(x+\mu) -
\phi(x)$. Here $x = (ai,aj,ak,al)$ where $a$ is the lattice spacing,
$i,j,k,l \in \lbrace 1,\ldots,N \rbrace$ are integers, and $x+\mu$ ($\mu
\in \lbrace 1,\ldots,4\rbrace$) denotes the lattice site which is the
nearest neighbor of $x$ in the forward $\mu$ direction. In the
classical theory, one can perform a point transformation $\phi(x)
\rightarrow \Phi(\phi(x))$, and the equations of motion in terms of the
new fields $\Phi(x)$ have the same content as the equations of motion in
terms of the old fields. However, the lattice action obtained from the
classical action expressed in terms of the fields $\Phi(x)$ will be
different from the lattice action obtained from the classical action
expressed in terms of the fields $\phi(x)$. The equivalence of the
theories at the continuum classical level rests on the chain rule for
differentiation, which is not valid on the lattice. It is not {\it a
priori\/} clear that the continuum QFT obtained at the second stage is
the same for the two different lattice actions.

In the great body of lattice field theory simulations, this possible
ambiguity is circumvented by appealing to the principle of universality.
The principle of universality maintains that, in the critical region,
the large distance behavior of lattice systems is mostly independent of
the small distance features, such as the detailed choice of the lattice
action.  In extrapolating to the continuum QFT, one reexpresses the
observables of interest in units of appropriate physical lengths. In the
continuum limit, as noted above, these finite physical units of length
become infinite as measured in terms of the lattice spacing. Therefore,
the specific features of the lattice action automatically become small
distance features as compared to anything of interest in the continuum
QFT.  Universality requires that these small distance features be
unimportant.

More precisely, all possible lattice models separate into classes. All
lattice field theories within a single universality class become the
same QFT in the continuum limit, but lattice models belonging to
different universality classes yield different continuum QFT's. From the
point of view of quantum field theory, the details of the lattice theory
do affect a small set of quantities that are nonuniversal in the
critical region, but which do not affect the physical content of the
continuum QFT.  On the other hand, in statistical mechanics, these
nonuniversal quantities may have physical content.  For example, in the
models considered in this article, it is found that the magnetization,
which is related to an expectation value of a field (Eq. \ref{magnet})
in one of the formulations of the $SO(2,1)$ model, has nonuniversal
behavior. The magnetization is relevant for the wavefunction
renormalization \cite{DDF:92a}, but apparently does not affect the
S matrix of the theory (which is trivial).

There are three criteria which determine a universality class. They are
(\cite{Gol:92} p.~80):
\begin{enumerate}
\item The spacetime dimension of the system;
\item The internal symmetry group of the system, and which
representation of this symmetry group is furnished by the fields in the
lattice action;
\item The nature of the critical point.
\end{enumerate}
These three criteria alone still leave one with the freedom to use many
different lattice actions to simulate a given theory. Although the above
criteria are widely known, additional criteria may be
needed to determine a universality class uniquely. {\it cf\/}
\cite{Kad:75} p.~18.

It is now seen how universality leads to a resolution of the ambiguity
inherent in formulating a lattice field theory. To ensure that the
lattice field theories under consideration go to the same continuum
limit, the lattice actions must have the same spacetime dimension,
have the same internal symmetry group, and be constructed out of
fields belonging to the same group representation. It is
{\em not\/} immediately clear that different lattice actions (such as
those obtained from the fields $\phi(x)$ and $\Phi(x)$ in the discussion
above), which satisfy criteria 1) and 2) will satisfy the third
 as well. It is commonly assumed that the first two criteria are
sufficient. This is occasionally tested when numerical studies are
carried out. This assumption is often made for scalar fields in four
spacetime dimensions, and is buttressed by the strong evidence
existing\cite{Cal:88} for the triviality of $\lambda\Phi^4$ theory.
However, for other scalar field theories, as yet unknown fixed points
may exist.

The principle of universality appears most natural in the framework of
the renormalization group (RG). A good introduction to the subject may
be found in \cite{Gol:92},\cite{Bel:91},\cite{Zin:89} or
\cite{Ma:73} while a detailed review is given in \cite{WK:74} and
\cite{Wil:75}. The following is a brief overview of the key ideas.
As noted previously, the continuum limit is extracted from the features
of the lattice theory with length scales much larger than the lattice
spacing. The RG provides a means of studying these features by gradually
integrating the high momentum degrees of freedom out of the lattice
version of the partition function.

A renormalization group transformation consists of the following steps:
\begin{enumerate}

\item Begin with an action defined at the scale of some cutoff
$\Lambda$ (on the lattice $\Lambda\sim{1\over a}$). This action should
be {\em the most general possible action consistent with the symmetries of
the theory\/};

\item Integrate modes with momenta in the range ${\Lambda\over s}\le k
\le\Lambda$, out of the functional integral. Here $s$ is some scale
factor greater than one;

\item Rescale all the remaining momenta by $k\rightarrow sk$ so as to
restore the original cutoff $\Lambda$;

\item Redefine the dummy field variables of integration in the partition
function by $\tilde\phi'(k) = \alpha(s)\tilde\phi(k)$, where $\alpha(s)$
is the function required to restore the kinetic term in the action
(${1\over 2}k^2\tilde\phi(k)^2$) to its original normalization.

\end{enumerate}
The effective action obtained after an RG transformation has the same
form as the original action, but the coupling parameters are modified.
RG transformations thus generate a flow in the space of coupling
parameters. In the cases of interest, this flow terminates at a fixed
point, at which the effective action is RG invariant.

If the parameters in the original action have their critical values and
the correlation length $\xi$ is infinite, then $\xi$ remains infinite
under RG transformations. The flows that leave $\xi$ infinite generate a
subspace of the coupling parameter space known as the {\em critical
surface}. It is assumed that every point on the critical surface flows
to some fixed point. The theory at this fixed point has the same long
distance behavior as any critical theory connected to it by RG
transformations. It should be emphasized that the RG transformation
changes the natural unit of length (the lattice spacing). The long
distance behavior is only the same in both cases {\em when the same unit
of length is used\/}. Therefore, the long distance behavior of the
original critical theory can be deduced by studying the theory at the
fixed point.

The origin of universality now becomes clear. The critical surface
breaks up into domains, such that all the points in a domain flow to the
same fixed point. Each domain defines a universality class. Within a
given universality class, the details of the lattice action of some
starting critical theory are not important because all theories have the
same long distance behavior as the fixed point theory.

As will be shown below, the long distance behavior for a theory which is
slightly off the critical surface, is the same as the behavior of the
theory at some point in the neighborhood of the fixed point. For
simplicity, the analysis of RG transformations in the neighborhood of
the fixed point will be restriced to the case of a single coupling
parameter. Under a general RG transformation,
\begin{equation}
K' = R(s,K)
\end{equation}
the function $R(s,K)$ is a smooth function of the coupling parameter $K$
for a finite rescaling of the cutoff by $s$. If $K$ is sufficiently
close to the fixed point, where the coupling parameter has its fixed
point value $K^*$, the general RG transformation is well approximated by
the linear part of its Taylor expansion:
$$
K' = K^* + \left.{\partial R(s,K) \over \partial K}
\right\vert_{K=K^*}\!\!\!\!\! (K - K^*)\,
\equiv\, K^* + \Lambda(s)(K - K^*) \, .
$$
Considering two successive RG transformations with scale parameters
$s_1$ and $s_2$, one finds
$$
K^{\prime\prime} = K^* + \Lambda(s_1)\Lambda(s_2)(K - K^*)
                 = K^* + \Lambda(s_1s_2)(K - K^*)\, .
$$
This is just  the semi-group property of the RG. The only possible
functional form for $\Lambda(s)$ is
\begin{equation}
\label{scaling}
\Lambda(s) = s^y\, .
\end{equation}
It is now seen that the behavior of the coupling constant $K$ near the
fixed point under RG transformations is determined by the exponent $y$.
If $y > 0$, the coupling parameter is enhanced under an RG
transformation, and $K$ is called a {\em relevant\/} coupling. If $y <
0$, it is suppressed, and it is called an {\em irrelevant\/} coupling.
If $y = 0$, the coupling remains constant under RG transformations (at
least in the linear approximation) and it is then called {\em
marginal\/}. In the general case of an infinite set of couplings $K_i$,
$\Lambda(s)$ becomes an infinite-dimensional matrix. The coupling
parameters no longer transform as simply as in Eq. (\ref{scaling}).
Instead, the {\em eigenvalues\/} of $\Lambda(s)$ transform in this
simple manner. These eigenvalues are usually called {\em scaling
fields\/}. The coupling parameters, however, are linear combinations of
the scaling fields. The operators multiplying the scaling fields in the
action are also called relevant, irrelevant, or marginal, according to
the behavior of the associated scaling fields.  Terms multiplying
irrelevant scaling fields in the effective action become small as
successive RG transformations drive the system to the fixed point.

If the system is initially slightly off the critical surface, the
direction of the RG flow of the relevant scaling fields will be away
from the fixed point (and hence the critical surface), while the
direction of flow of the irrelevant scaling fields will be towards the
fixed point. The relevant scaling fields must vanish on the critical
surface.  Otherwise, RG transformations will take the system out of the
critical surface. Therefore, from a starting point sufficiently close to
the critical surface, RG transformations first drive the system to the
neighborhood of the fixed point, because the irrelevant operators
dominate. The long distance behavior of the original critical region is
thus seen to be the same as the long distance behavior in a neighborhood
of the fixed point, since flow trajectories connect the two regions.
Eventually, the relevant scaling fields will drive the system away from
the neighborhood of the fixed point.

The RG also explains why the critical exponents of certain quantities
are universal. Their critical behavior can be analyzed in terms of the
behavior near the fixed point. Such quantities themselves obey
transformation laws similar to those of the coupling parameters:
\begin{equation}
A^\prime = s^{y_A}A\, .
\end{equation}
One such quantity is the correlation length, whose scaling behavior is
$$
\xi(K_1,0,\ldots,L_1,L_2,\ldots) \rightarrow
\xi'(K'_1,0,\ldots,L'_1,L'_2,\ldots) =
s^{-1}\xi(s^{y_1}K_1,0,\ldots,s^{x_1}L_1,s^{x_2}L_2,\ldots)\, ,
$$
where $K_i$ are the relevant scaling fields and $L_i$ are the irrelevant
scaling fields. The system is considered initially to have all the
scaling fields except $K_1$ set to zero, and $K_1\ll 1$, since the
initial point is very close to the critical surface.  Performing a
transformation with $s = K_1^{-{1\over y_1}}$, one gets
$$
\xi'(K'_i,L'_i) =
K^{1\over y_1}\xi(1,0,\ldots,K_1^{-{x_1\over y_1}}L_1,
                        K_1^{-{x_2\over y_1}}L_2,\ldots)\, .
$$
Since $K_1^{-{1\over y_1}}\gg 1$, and $x_i < 0$ for all $i$, it follows
that
$$
\xi'(K'_i,L'_i) \approx K_1^{-{1\over y_1}}\xi(1,0,0,\ldots,0,0,\ldots)
              \sim K_1^{-{1\over y_1}}\, .
$$
This means that the critical exponent for the correlation length with
respect to the coupling $K_1$ is $-{1\over y_1}$. The critical exponents
of $\xi$ with respect to any other relevant coupling parameter can be
obtained by starting with {\em that} coupling as the only non-zero relevant
coupling parameter. The reason that the critical exponents are
universal now emerges. The values of the critical exponents are
generically determined by the eigenvalues of the linearized RG near the
fixed point in the given universality class, and not by the values of
the coupling parameters themselves. Any system in a given universality
class will therefore have the same critical exponents with respect to
given relevant coupling parameters (such as the temperature, the
external field, etc.).

%%% -- End of section 1a (second section in paper)

%
% Section 2, Nonlinear Sigma Models.
%

\section{Nonlinear Sigma Models}
\label{models}

Numerical results for the nonlinear sigma models with noncompact
configuration spaces demonstrate the need for a closer look at
universality. Nonlinear sigma models are a class of scalar field
theories that can be obtained by imposing a constraint among a set of
free massless scalar fields\cite{Zin:89}.  The models are constructed to
be invariant under the {\em global} transformations of some
finite-dimensional Lie group, $G$. The configuration space in which the
fields live is generally a coset space of $G$.  For the {\em compact}
$SO(2)$ and $SO(3)$ sigma models, the configuration spaces are the
circle and the 2-sphere respectively.  These sigma models are analogous
to spins systems in statistical mechanics.  If $G$ is a noncompact
group, then the configuration space may also be noncompact.  This
article will be concerned only with Euclidean models, invariant under
$SO(n,m)$ groups.

Consider a set of $k+1=m+n$ massless free scalar fields whose Euclidean
action is invariant under transformations of $G$.
\begin{equation}
S[\varphi] = \frac{1}{2} \int d^{4}x \;
\eta_{ij} \; \partial_{\mu}\varphi^{i}
\partial_{\mu}\varphi^{j},\hspace{1cm}i,j\in \{1,2,\ldots,k+1\}.
\label{freeaction}
\end{equation}
The fields $\varphi^{i}$ are cartesian coordinates for the internal
space associated with each space-time point. In this coordinate system,
$\eta_{ij}$ is diag$(\underbrace{1, 1, \ldots}_{n}, \underbrace{-1, -1,
\ldots}_{m})$.

If the fields at every point are required to satisfy the constraint,
\begin{equation}
-\eta_{ij} \; \varphi^{i} \varphi^{j} = \mu^2,
\label{constraint}
\end{equation}
then the theory described by (\ref{freeaction}) together with
(\ref{constraint}) is known as a nonlinear sigma model. The
constraint merely requires that the length of the field vector (in the
metric of the configuration space) be a constant. Since both
$S[\varphi]$ and the constraint are invariant under the global action of
$G$, the sigma model clearly possesses the same symmetry. Different
groups describe different theories, and the symmetry is generally used
to distinguish among them.

The fields are constrained to a $k$-dimensional curved surface, $C$, in
the $(k+1)$-dimensional flat space. This surface is a space of constant
curvature and is generally the manifold of a coset space of $G$. The
space $C$ may be noncompact, if $G$ is a noncompact group. The curvature
of the configuration space introduces interactions into
the theory. The nature of these interactions is most easily seen if the
constraint is used to eliminate one of the fields from the theory.
After eliminating $\varphi^{k+1}$, the action can be expressed in terms
of the remaining fields and $G_{ab}(\varphi)$, the induced metric on the
constraint surface.
\begin{equation}
S[\varphi] = \frac{1}{2} \int d^{4}x \;
G_{ab}(\varphi) \partial_{\mu} \varphi^{a} \partial_{\mu}\varphi^{b},
\hspace{1cm} a,b\in \{1,2,\ldots,k\},
\label{sigmaction1}
\end{equation}
\begin{displaymath}
G_{ab}(\varphi) = \eta_{ab} - \frac{\eta_{ac}\eta_{bd}\varphi^{c}\varphi^{d}}
{\mu^{2} + \eta_{rs}\varphi^{r}\varphi^{s}},
\hspace{1cm} a,b,c,d,r,s \in \{1,2,\ldots,k\}.
\end{displaymath}
Expanding $G_{ab}(\varphi)$ around $\varphi^{a}=0$ gives
\begin{eqnarray}
S[\varphi] & = & \frac{1}{2} \int d^{4}x \;
\left[ \eta_{ab} \partial_{\mu}\varphi^{a}\partial_{\mu}\varphi^{b}
- \mu^{-2} \; \eta_{ac}\eta_{bd}\varphi^{c}\varphi^{d}
\partial_{\mu}\varphi^{a}\partial_{\mu}\varphi^{b}  \right. \nonumber \\
&  & \qquad \qquad \;\; + \left. \mu^{-4} \; \eta_{ac}\eta_{bd}\eta_{rs}
\varphi^{c}\varphi^{d}\varphi^{r}\varphi^{s}
\partial_{\mu}\varphi^{a}\partial_{\mu}\varphi^{b} - \ldots \right]
\label{expansion}
\end{eqnarray}

It is clear from (\ref{expansion}) that $\mu^{-2}$ plays the role of a
coupling constant.  It is also seen that the interactions involve field
derivatives and that the action is not polynomial in nature. In four
dimensions, $\mu$ has dimensions of mass and the coupling constant has
dimensions of $[{\rm mass}]^{-2}$. Well known power counting arguments
indicate that theories with couplings having negative mass dimension are
not perturbatively renormalizable \cite{IZ:80}.  In addition, naive
scaling arguments indicate that that $1/\mu^{2}$ will be irrelevant
under the renormalization group transformation described in Sec.
\ref{univ}. For this reason, it is generally believed that the nonlinear
sigma models in four dimensions are all in the same universality class
as the free field theory (they are trivial).

For purposes of numerical simulation, it is convenient to render the
fields dimensionless by the rescaling $\phi^{a}=\varphi^{a}/\mu$. In
terms of these dimensionless fields, the sigma model action becomes
\begin{equation}
S[\phi] = \frac{\mu^{2}}{2} \int d^{4}x \hspace{2mm} G_{ab}(\phi)
\partial_{\mu}\phi^{a} \partial_{\mu}\phi^{b},\hspace{1cm}a,b
\in \{1,2,\ldots,k\},
\label{sigmaction2}
\end{equation}
where
\begin{displaymath}
G_{ab}(\phi) = \eta_{ab} - \frac{\eta_{ac}\eta_{ad}\phi^{c}\phi^{d}}
{1 + \eta_{rs}\phi^{r}\phi^{s}},\hspace{1cm}
a,b,c,d,r,s \in \{1,2,\ldots,k\}.
\end{displaymath}

When spacetime is taken to have periodic boundary conditions, any
constant field is a solution of the classical dynamical equations which
minimizes the Hamiltonian. The existence of a continuous global symmetry
dictates that sigma models have a k-parameter family of ground state
solutions. Because of this, the models display the phenomenon of {\em
spontaneous symmetry breaking}, in which only one of the possibilities
is realized as the physical vacuum.

Euclideanized nonlinear sigma models with compact symmetry groups have
been widely studied as statistical mechanical spin systems, and the
universality classes of these models appear to be distinguishable by the
standard rules \cite{Ami:84}. As will be shown in this article, the
situation is {\em different\/} for sigma models with noncompact symmetry
groups. The $SO(1,1)$ and $SO(2,1)$ models are the simplest sigma models
of this type. Both of these exhibit behavior that would normally be
considered nonuniversal. The $SO(1,1)$ model is particularly
illustrative, since it is classically related to the free theory by a
transformation of field variables. Different discretizations, however,
lead to different lattice theories, only one of which is {\em simply\/}
related to the continuum free theory.

%
% Section 3, The SO(1,1) Model.
%

\section{The $SO(1,1)$ Model}
\label{o11}

\begin{figure}[htb]
\epsfxsize=5in
\centerline{\epsffile{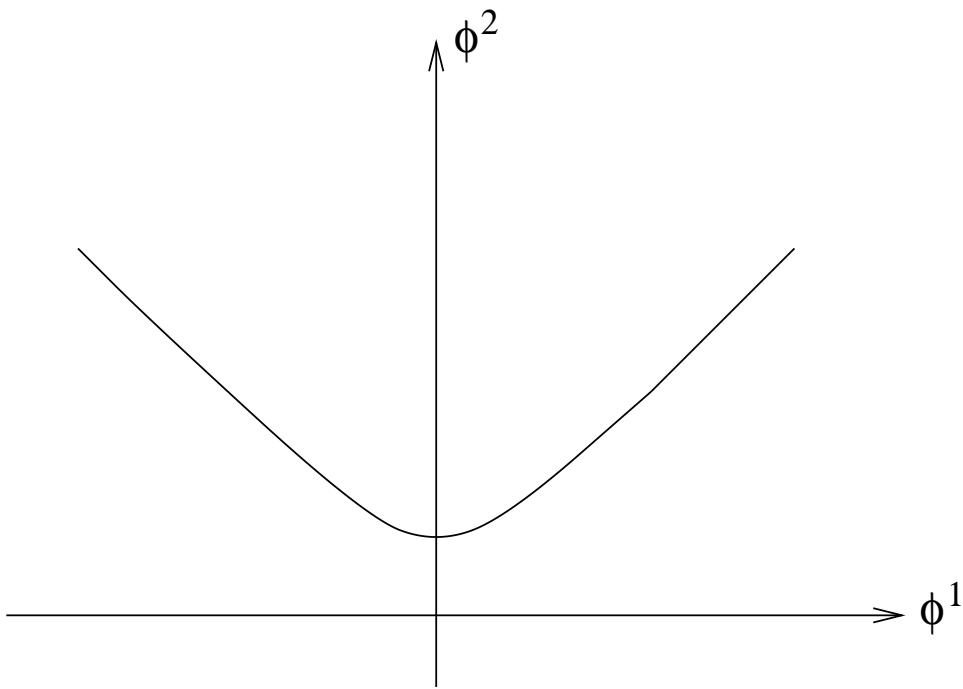}}
\caption{Configuration Space for the $SO(1,1)$ Sigma Model.}
\label{o11man}
\end{figure}

The $SO(1,1)$ Sigma Model is defined by (\ref{sigmaction1}) with
$(k+1)=2$ and $\eta_{ij}=$ diag$(1,-1)$. In terms of dimensionless
fields, the action is
\begin{equation}
S =  \frac{\mu^{2}}{2} \int d^{4}x \hspace{2mm}
(\partial_{\mu}\phi^{1}\partial_{\mu}\phi^{1} -
 \partial_{\mu}\phi^{2}\partial_{\mu}\phi^{2}),
\label{o11action1}
\end{equation}
subject to the constraint that $(\phi^2)^{2} - (\phi^1)^{2} = 1$.  This
constraint surface is disconnected, so the theory will be restricted to
the connected component defined by $\phi^{2} > 0$.  The internal space
shown in Fig. \ref{o11man} is the coset space $SO(1,1)/(Z_{2} \times
Z_{2})$. It is topologically equivalent to {\bf R} and is intrinsically
flat. Group transformations leave the surface invariant.

%
%       Figure 1 should be here.
%

The $SO(1,1)$ Sigma Model defined in this way is classically
equivalent to the theory of a massless free scalar field.  Let
$\phi^{1}=\sinh{s}$ and $\phi^{2}=\cosh{s}$ in (\ref{o11action1}). The
action in terms of the field $s$ is
\begin{equation}
S = \frac{\mu^{2}}{2} \int d^{4}x
\partial_{\mu} s \partial_{\mu} s,
\label{o11saction}
\end{equation}
which is just the massless free theory. The variable $s$ measures the
arc length from the $\phi^{2}$ axis along the hyperbola in Fig.
\ref{o11man}.

For numerical simulations of the quantum theory, the $SO(1,1)$ Sigma
Model must be defined on a four-dimensional hypercubical lattice with
periodic boundary conditions. The lattice consists of $N^{4}$ points (or
sites), $N$ along each spacetime direction. The sites may be connected
by imaginary links, which join each site to its nearest neighbors.  The
distance between lattice sites is denoted by $a$, and the length of one
side of the lattice is $L=Na$.  Lattice approximations to the continuum
action are generally obtained, as noted in Sec.\ref{univ}, by replacing
derivatives with lattice differences and integrals by sums.  When this
procedure is performed on the $SO(1,1)$ model, the action takes the form
\begin{equation}
S = \frac{\beta}{2} \sum \cal{L}(\phi).
\end{equation}
Here $\beta$ is $\mu^{2}a^{2}$---a dimensionless lattice version of the
coupling.

The exact form of $\cal{L}(\phi)$ depends on the details of the
differencing procedure (two of which will be discussed shortly), but
it generally contains interactions which tend to align the fields at
neighboring sites. In this respect, the lattice sigma models are
analogous to ferromagnetic spin systems in statistical mechanics. The
field values at each site play the role of magnetic spins, and $\beta$
plays the role of an inverse temperature. This analogy is more easily
visualized for the compact sigma models, but it is useful for the
noncompact models as well.

If the direction of symmetry breaking is chosen so that the average
value of $\phi^{1}$ is zero (this must be enforced by hand in
finite-lattice numerical
simulations), then the average value of $\phi^{2}$ is a natural order
parameter for this model. When $\beta$ is large,
$\langle\phi^{2}\rangle$ is near unity. This indicates that the fields
at most sites are nearly aligned (highly ordered) around $s = 0$. For
small $\beta$, $\langle \phi^{2} \rangle$ is large while
$\langle\phi^{1}\rangle$ remains zero. This indicates that the fields
fluctuate widely over the configuration space and that the system is
highly disordered.

Evidently, $\langle \phi^{2} \rangle$ is analogous to the magnetization
in ferromagnetic systems, but because the configuration space is not
compact, it behaves quite differently. In order to define a
magnetization that behaves in a familiar fashion, the quantity
\begin{equation}
\label{magnet}
M = \frac{1}{\langle \phi^{2} \rangle}
\end{equation}
will be designated the ``magnetization'' for this model.

\begin{figure}[hbt]
\epsfxsize=5in
\centerline{\epsffile{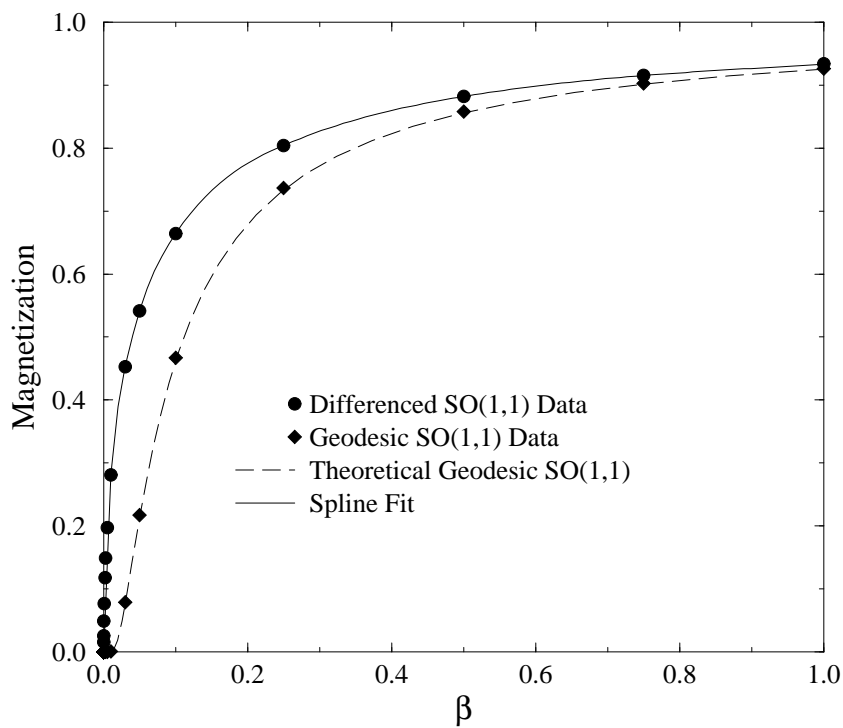}}
\caption{Magnetizations for the $SO(1,1)$ Sigma Model.}
\label{o11mag}
\end{figure}

Ferromagnetic systems generally have some characteristic temperature,
called the critical temperature, below which the system becomes
spontaneously aligned. Sigma models generally have a characteristic
value of $\beta$, denoted by $\beta_{c}$, {\em above} which the fields
begin to become aligned.  This is called the critical value of $\beta$,
or the critical point of the theory. The value of $\beta_{c}$ depends on
the lattice model and the dimension of spacetime, but for compact
models, it is generally nonzero. {\em For noncompact sigma models,
however, $\beta_{c}$ is always zero.} This is because the fields are
always aligned to some degree unless an infinite ``temperature''
($\beta=0$) induces fluctuations over the entirety of the noncompact
configuration space. For a finite temperature (finite $\beta$),
fluctuations will be finite and the magnetization will be nonzero. A
plot of the magnetization as a function of $\beta$ for two lattice
$SO(1,1)$ models in Fig.
\ref{o11mag} illustrates this.

The normal lattice approximation to the integrand of (\ref{o11saction})
is simply
\begin{equation}
(\partial_{\mu}s)^{2} \rightarrow \frac{1}{a^{2}} (\Delta_{\mu} s)^{2},
\hspace{5mm} \Delta_{\mu}s = s(x+\mu)-s(x),
\label{sintegrand}
\end{equation}
where the 4-vectors $\mu$ are given by
\begin{eqnarray}
1 = (a,0,0,0), & \hspace{1cm} & 2 = (0,a,0,0), \nonumber \\
3 = (0,0,a,0), & \hspace{1cm} & 4 = (0,0,0,a). \nonumber
\end{eqnarray}
Application of the same prescription to the integrand of
(\ref{o11action1}) gives
\begin{eqnarray}
(\partial_{\mu}\phi^{1})^{2} - (\partial_{\mu}\phi^{2})^{2}
& \rightarrow & a^{-2} \left\{ (\Delta_{\mu}\phi^{1})^{2}
- (\Delta_{\mu}\phi^{2})^{2} \right\}  \nonumber \\
& = & a^{-2} \{ [\sinh(s(x+\mu))-\sinh(s(x))]^{2} \nonumber \\
& & \quad \, - [\cosh(s(x+\mu))-\cosh(s(x))]^{2} \} \nonumber \\
& = & 4a^{-2} \sinh^{2}\left(\frac{\Delta_{\mu}s}{2} \right)
\nonumber \\
& = & a^{-2} \left\{ (\Delta_{\mu}s)^{2}
+ \frac{1}{12} (\Delta_{\mu}s)^{4} + \cdots  \right\}
\label{phiintegrand}
\end{eqnarray}

As $a \rightarrow 0$, the expressions (\ref{sintegrand}) and
(\ref{phiintegrand}) clearly become the same when the fields are smooth.
In four dimensions, however, the fields are not smooth.  One of the odd
properties of functional integrals is that the field configurations that
contribute significantly to the integral are generally neither
continuous nor differentiable (except in one dimension, where they are
continuous). This issue is discussed at length in \cite{DFG:91}, but
qualitatively correct behavior can be surmised from the rule of thumb
that configurations which contribute significantly to the functional
integral are those for which, $\langle S
\rangle \sim N^{d}$, where $d$ is the dimension of spacetime. Applying
this rule of thumb to the lattice action implied by (\ref{sintegrand})
gives
\[ \Delta_{\mu}s \sim a^{1-d/2}. \]
The same rule applied to (\ref{phiintegrand}) gives
\[ \Delta_{\mu}s \sim \ln(a) \]
as $a \rightarrow 0$. This causes the lattice theories defined by
(\ref{sintegrand}) and (\ref{phiintegrand}) to behave very differently,
despite the fact that they were obtained from the same classical action.

The numerical results\cite{Mye:92} shown in Fig. \ref{o11mag} clearly
confirm that (\ref{sintegrand}) and (\ref{phiintegrand}) produce
significantly different results for the critical exponent of the
magnetization. These numerical simulations were performed on a lattice
of $8^4$ sites; at this point, accurate error estimates are unavailable.

%
%       Figure 2 should be here.
%

In fact, the $\beta$ dependence of the magnetization corresponding to
(\ref{sintegrand}) can be calculated analytically and is found
\cite{DDF:92b} to behave as $\exp(-\frac{{\rm const.}}{\beta})$,
which does not allow a definition of a critical exponent at all. The
numerical results for (\ref{phiintegrand}), however, indicate that the
magnetization
behaves approximately as $\beta^{1.4}$ near the critical point. There is
a significant {\em qualitative} difference between these two lattice
models.

The radical difference in the behavior of the magnetizations for these
two theories raises the possibility that the two models may belong to
different universality classes; the model defined by
(\ref{phiintegrand}) may not become a free continuum field theory. The
numerical simulations performed to date\cite{Mye:92}, do not settle this
question definitively. The same simulations, however, {\em do\/} show
some differences in the critical exponents for the specific heat, $c_L$,
of the two models (this result is further discussed in \cite{DDF:92b}).
For the model with the action (\ref{phiintegrand}),
$c_L\sim\beta^{-1.88}$, while for (\ref{sintegrand}),
$c_L\sim\beta^{-2.00}$. This difference, while not so drastic as the one
for the magnetizations, appears to be another example of nonuniversal
behavior.

%%% -- End of third section
%
% Section 4, The SO(2,1) Model.
%

\section{The $SO(2,1)$ Model}
\label{o21}

\begin{figure}[htb]
\epsfxsize=5in
\centerline{\epsffile{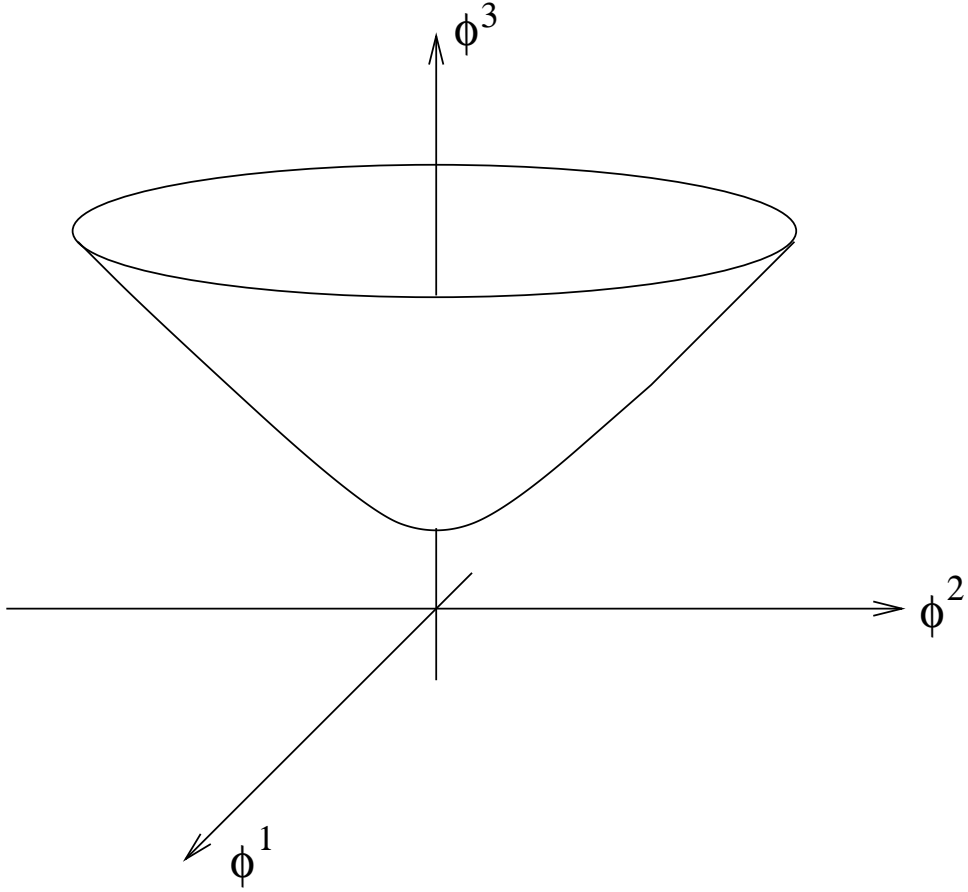}}
\caption{Configuration Space for the $SO(2,1)$ Sigma Model.}
\label{o21man}
\end{figure}

The $SO(2,1)$ Sigma Model is defined by (\ref{sigmaction1}) with
$(k+1)=3$ and $\eta_{ij}=$ diag$(1,1,-1)$. In terms of dimensionless
field variables,
\begin{equation}
S =  \frac{\mu^{2}}{2} \int d^{4}x
\left( \partial_{\mu}\phi^{1}\partial_{\mu}\phi^{1} +
\partial_{\mu}\phi^{2}\partial_{\mu}\phi^{2} -
\partial_{\mu}\phi^{3}\partial_{\mu}\phi^{3} \right),
\label{o21action}
\end{equation}
with the constraint that
%\begin{displaymath}
$(\phi^3)^{2} - (\phi^2)^{2} - (\phi^1)^{2} = 1$.
%\end{displaymath}
%
Again, the connected sheet $\phi^{3} > 0$ is chosen.
The resulting configuration space, shown in Fig.
\ref{o21man}, is $SO(2,1)/(SO(2) \times Z_{2})$. This is topologically
equivalent to ${\bf R}^{2}$ but is not flat.

Substitution of finite differences into (\ref{o21action}) produces one
possible lattice action for the $SO(2,1)$ model. Another can be obtained
by means of {\em geodesic} differencing. Let $l$ be the geodesic length
between two field values on the hyperboloid shown in Fig.
\ref{o21man}.  Geodesic differencing prescribes\footnote{Reasons for
using this prescription are discussed in \cite{DFG:91}.} that the
lattice Lagrangian for the $SO(2,1)$ model be $l^{2}/a^{2}$. (Note that
geodesic differencing of the $SO(1,1)$ model yields (4.6).)

%
%       Figure 3 should be here.
%

If $s$ is the geodesic length along the internal space from the point
$(0,0,1)$ to the point $(\phi^{1},\phi^{2},\phi^{3})$, and $\theta$ is
the angle measured from the $\phi^{1}$ axis, then the geodesic lattice
action for the $SO(2,1)$ model is \cite{DDF:92b}

\begin{eqnarray}
S & = & \frac{\beta}{2} \sum_{x,\mu} \left\{
\cosh^{-1}\left[ \cosh[s(x+\mu)-s(x)]\cos^{2}
\left( \frac{\theta(x+\mu)-\theta(x)}{2} \right) \right. \right.
\nonumber \\
&  & \left. \left. \qquad \quad + \cosh[s(x+\mu)+s(x)]\sin^{2}
\left( \frac{\theta(x+\mu)-\theta(x)}{2} \right) \right] \right\}^{2}.
\label{slataction}
\end{eqnarray}
Again, $\beta=\mu^{2}a^{2}$ may be interpreted as the reciprocal of a
bare lattice coupling, analogous to an inverse temperature in a
statistical mechanical model. In a fashion similar to the $SO(1,1)$
model, the magnetization is defined to be $\langle\phi^{3}\rangle^{-1}$
(with $\langle \phi^{1} \rangle = \langle \phi^{2} \rangle = 0$).

\begin{figure}[hbt]
\epsfxsize=5in
\centerline{\epsffile{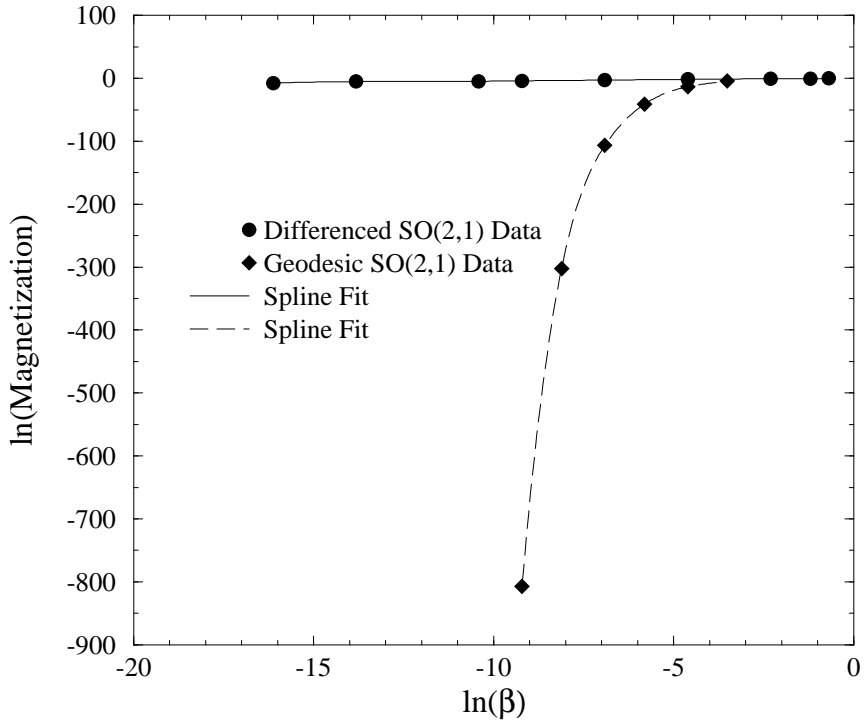}}
\caption{Log Plot of Magnetizations for the $SO(2,1)$ Sigma Model.}
\label{o21mag}
\end{figure}

Plots of the magnetizations for the two lattice models described above
are shown in Fig. \ref{o21mag}.  For the geodesic model, the $\beta$
dependence of the magnetization near the critical point is analytically
soluble \cite{DDF:92b} and has exponential behavior similar to that
displayed by the geodesically differenced $SO(1,1)$ model. The model
obtained by differencing (\ref{o21action}), however, has a magnetization
that displays the familiar power-law behavior near $\beta=0$. For a
lattice of $10^4$ sites, the magnetization behaves approximately as
$\beta^{0.5}$ (error estimates for this value are presently
unavailable). Again, there is a striking {\em qualitative} difference
between the two lattice models.

%
%       Figure 4 should be here.
%

Extensive numerical simulations of the geodesically-differenced
$SO(2,1)$ model indicate clearly that this theory is trivial in the
continuum limit \cite{DDF:92a,DDF:92b}. Somewhat less thorough, but
still convincing numerical studies for the naively-differenced theory
\cite{DDF:88} suggest the that this theory is also trivial. In this
respect, expectations from the ideas of universality appear to be
fulfilled.  Nevertheless, the difference between magnetizations of the
two lattice models remains to be explained. If there were a noncompact
{\em statistical mechanical model\/}, the difference would certainly be
physically significant.

%%% -- End of fourth section
%
% Section 5, Conclusions.
%

\section{Conclusions}
\label{concl}

The noncompact nonlinear sigma models discussed here clearly display
behavior that would ordinarily be considered a violation of
universality.  One possible explanation for this behavior is that the
different lattice theories are associated with different universality
classes. This explanation, however, is difficult to reconcile with the
renormalization group arguments about irrelevant operators, which
suggest that the different lattice theories should have the same
continuum limit.  It was mentioned in Secs. \ref{o11} and \ref{o21}
that the numerical evidence supports (to varying degrees) the view that,
as quantum field theories, these noncompact models are free (trivial).
This view is tenuous for the $O(1,1)$ model (the data is also consistent
with a nontrivial continuum limit), but is well supported by
numerical results for the $O(2,1)$ theory. If it is true, it means that,
under the action of the renormalization group, the different lattice
theories flow to the Gaussian fixed point.

In such a case, it must be that the critical exponent for the
magnetization is one of the quantities in the theory that does not
behave in a universal fashion. This opposes a large body of evidence,
but much of that has been obtained in the context of models with compact
configuration spaces.  To understand this nonuniversal behavior, examine
the standard argument why the magnetization should have universal power
law behavior near the critical point:

\begin{enumerate}

\item \label{rgrelation}
        The critical lattice theory is connected by renormalization
        group transformations to another theory, which is at the fixed point.
        The theory at the fixed point has the same long-distance
        behavior as the original theory.

\item \label{couplingscale}
         Near a fixed point of the renormalization group transformation,
        the coupling constant $\beta$ will scale by some power of $s$,
        $\beta \rightarrow \beta' = s^{y}\beta$.

\item \label{fieldscale}
        The field will also scale by
        $s$ to some power, $\phi \rightarrow s^{-d_{\phi}}\phi$. Since
        the magnetization is generally just $\langle \phi \rangle$,
        it picks up an overall factor of $s^{-d_{\phi}}$.

\item \label{magscale}
        The above behavior leads to the relationship $M'(\beta')=
        s^{-d_{\phi}}M(s^{y}\beta)$. This relationship implies
        $M(\beta) \sim \beta^{\frac{d_{\phi}}{y}}$.

\end{enumerate}

It is the third of the above statements that is false when applied to the
magnetization of the lattice free field and the geodesic $O(2,1)$
model. In those cases, the magnetization is not just the average value
of the field; it does not scale in the same fashion as the field in the
kinetic term of the action. This means that the factor of
$s^{-d_{\phi}}$ in front of the magnetization is not correct in this
case, and the argument for a universal exponent fails. This should not
affect the universality of other critical exponents in the theory,
except perhaps those that are related to the exponent for the
magnetization.

The fact that the configuration spaces for these models are not compact
plays a major role in the observed differences in critical behavior. In
the action (\ref{phiintegrand}), the hyperbolic sine differs greatly
from $\Delta_{\mu}s$ over the range of field values that contribute
significantly to the integral (recall that the fields which contribute
to the functional integral are generally discontinuous \cite{DFG:91}).
This causes the fields to spread over the configuration space at an
exponential rate as $\beta \rightarrow 0$, resulting in an average
field which scales exponentially with $\beta$, instead of with the usual
power law. This type of scaling would be unlikely if the configuration
space were compact, simply because a compact configuration space places
a limit on the size of the fields.

It may seem unusual that the local scaling of the fields can affect
behavior that is normally considered to be due to long range
correlations. While it is the long range correlations that cause the
fields to align, the $value$ of the order parameter is influenced by the
local scaling (with $\beta$) properties of the field. For the sigma
models discussed above, the local behavior is sufficiently different to
cause qualitative differences in the behavior of the magnetization near
the critical point.

In this sense, the behavior of the magnetization is like that of the
critical coupling (or critical temperature). The value of the critical
coupling is known to depend strongly on the local details of the lattice
model; it is not a universal quantity. The conclusion of this article is
that, for some noncompact models, the same is true of the magnetization.
In field theory, this magnetization is not an observable quantity, and
its nonuniversal behavior should not affect the continuum theory.  As
noted in Sec. \ref{univ}, however, it is significant from the
statistical mechanical viewpoint.

Regarding conclusions about the triviality of these models, there
remains uncertainty about the continuum limit of the model defined by
(\ref{phiintegrand}).  More complete numerical simulations of this model
are necessary before its continuum limit can be cited with conviction.
In addition, it would be helpful to obtain an analytical approximation
to this theory for small values of $\beta$, similar to the one that
supports the numerical data for the $O(2,1)$ model
\cite{DDF:92b}. These tasks remain for future research.

%%% -- End of fifth section

% Acknowledgments.

\acknowledgments

The authors wish to thank Michael Mandelberg for helpful conversations,
and See Kit Foong for providing comments and numerical data. They are
grateful to Eric Myers for sharing his numerical results for the
$SO(1,1)$ model. As always, Bryce DeWitt started this mess, and should
be mentioned here also.

%% Just before electronic submission comment out the \bibliography and
%% the \bibliographystyle commands, and insert the universality.bbl file

% -- \input figs

\end{document}